\documentclass[twocolumn,showpacs,preprintnumbers,amsmath,amssymb,pre]{revtex4}
\usepackage{graphicx}
\usepackage{dcolumn}
\usepackage{bm}

\begin{document}

\title{\bf Emergence of quasi-units in the one dimensional Zhang model }
\author{Tridib Sadhu}
\email{tridib@tifr.res.in}
\author{Deepak Dhar}
\email{ddhar@theory.tifr.res.in}
\affiliation{Department of Theoretical Physics, \\  
Tata Institute of Fundamental Research, \\
Homi Bhaba Road, Mumbai 400005, India.}

\date{\today}
\begin{abstract} We study the Zhang model of sandpile on a one 
dimensional chain of length $L$, where a random amount of energy is 
added at a randomly chosen site at each time step. We show that in spite 
of this randomness in the input energy, the probability distribution 
function of energy at a site in the steady state is sharply peaked, and 
the width of the peak decreases as $ {L}^{-1/2}$ for large $L$. We 
discuss how the energy added at one time is distributed among 
different sites by topplings with time. We relate this distribution to 
the time-dependent probability distribution of the position of a 
marked grain in the 
one dimensional Abelian model with discrete heights.  We argue that in 
the large $L$ limit, the variance of energy at site $x$ has a scaling 
form $L^{-1}g(x/L)$, where $g(\xi)$ varies as $\log(1/\xi)$ for 
small $\xi$, which agrees very well with the results from 
numerical simulations.  
\end{abstract}

\maketitle

\section{Introduction}

After the pioneering work of Bak, Tang and Wiesenfeld (BTW) in 1987 
\cite{btw}, many different models for self-organized criticality have 
been studied in different contexts; for review see 
[2-5].  Of these, models in the general class known as Abelian distributed 
processors  have been studied a lot, as they share an Abelian 
property that makes their theoretical study simpler \cite{deepak}. The 
original sandpile model of Bak \textit{et al.} \cite{btw}, the Eulerian walkers 
model \cite{euler}, and the Manna model \cite{manna} are all members 
of 
this class.  Models which do not have the Abelian property have been 
studied mostly by numerical simulations.  In this paper, we 
discuss the Zhang model \cite{zhang}, which does not have the Abelian 
property.

In the Zhang model, the amount of energy added at a randomly chosen site 
at each time step is not fixed, but random.  In spite of 
this, the model in one dimension has the remarkable property that the 
energy at a site in the steady state has a very sharply peaked 
distribution in which the width of the peak is much less than the 
spread in the input amount per time step, and the width decreases with 
increasing system size $L$. This behavior was noticed by Zhang 
using numerical simulations in one and two dimension \cite{zhang}, and he called it 
the `emergence of quasi-units' in the 
steady state of the model. He argued that for large systems, the 
behavior would be same as in the discrete model. Recently, A. Fey 
\textit{et al.} \cite{redig} have proved that in one dimension, the 
variance of energy does go to zero as the length of the chain $L$ goes 
to infinity, but they did 
not study how fast  it decreases with $L$.

In this paper, we study this emergence of `quasi-units' in one 
dimensional Zhang sandpile by looking at how the added energy is redistributed 
among different sites in the avalanche process. We show that the 
distribution function of the fraction of added energy at a site $x'$ 
reaching a site $x$
after $t$ time steps following the addition is exactly equal to the 
probability distribution that a marked grain in the 
one-dimensional height type BTW model added at site $x'$ reaches site $x$ in 
time $t$. The latter problem has been 
studied recently \cite{punyabrata}. We use this to show that the 
variance of energy asymptotically vanishes as $ 1/L$.  We also 
discuss the spatial dependence of the variance along the system length.  In 
the large $L$ limit, the variance at site $x$ has a scaling form 
$L^{-1}g(x/L)$. We determine an approximate form of the scaling 
function $g(\xi)$, which agrees very well with the results of our 
numerical simulations.

There have been other studies of the Zhang model earlier. 
Blanchard \textit{et al.} \cite{blanchard} have studied the steady state of the model, and found 
that the 
distribution of energies even 
for the two site problem is very complicated, and has a multi-fractal 
character. In two dimensions, the distribution of energy seems 
to sharpen for larger $L$, but the rate of decrease of the width is very 
slow \cite{janoci}. Most other studies have dealt with the question as to whether the critical exponents of the avalanche distribution in this 
model are the same as in the discrete Abelian model \cite{lubeck, milshetein}
. A. Fey \textit{et al.}'s results imply that the asymptotic 
behavior of the avalanche distribution in one dimension is indeed the same as in 
the discrete case, but the situation in higher dimension remains unclear 
\cite{vespignani, guilera}.

The plan of the paper is as follows. In Section II, we define the model 
precisely.  In Section III, we show that the calculation of the way the 
energy added at a site is distributed among different sites by 
toppling is same as the calculation of the time-dependent probability 
distribution of the position  of a marked grain in the 
discrete Abelian sandpile model. This correspondence is used in Section 
IV to determine the qualitative dependence of the variance of the energy 
variable at a site on its position $x$, and on the system size $L$. We 
propose a simple extrapolation form that incorporates this dependence. 
We check our theoretical arguments with numerical simulations in Section V.  
 Section VI contains a summary and concluding remarks. A detailed calculation
of the solution of an equation, required in Section IV, is added as an Appendix.

\section{Definition and preliminaries}

We consider our model on a linear chain of size $L$. The sites are 
labelled by integers $1$ to $L$ and a real continuous energy variable is 
assigned to each site. Let $E(x,t)$ be the energy variable at site $x$ 
at the {\it end} of the time-step $t$. We define a threshold 
energy 
value $E_c$, same for each site, such that sites with $E(x,t) \ge E_c $ are 
called unstable, and those with $E(x,t) < E_c$ are called stable.  
Starting from a configuration where all sites are stable, the 
dynamics is defined as follows.

(i) The system is driven by adding a random amount of energy at the 
\textit{beginning} of every   time-step at a randomly chosen site. Let the amount 
of energy added 
at time $t$ be $\Delta_t$.  We will assume that all $\Delta$'s are 
independent, identically distributed random variables, each picked 
randomly from an uniform interval $1-\epsilon \le \Delta_t \le 
1+\epsilon$. Let the site of addition chosen at time $t$ be denoted by 
$a_t$.

(ii) We make a list of all sites whose energy exceeds or becomes equal to 
the critical value $E_c$.  All these sites are relaxed in parallel by 
topplings. In a toppling, the energy of the site is equally distributed to its 
two neighbors and the energy at that site is reset to zero. If there is 
toppling at a boundary site, half of the energy at that site before 
toppling is lost. 

(iii) We iterate Step (ii) until all topplings stop. This completes one 
time step. 

This is the slow driving limit, and we assume that all avalanche 
activity stops before the next addition event. In this limit, the model 
is characterized by two parameters $\epsilon$ and $E_c$. In the limit 
$\epsilon =0$, and $1 < E_c \le 2$, the model reduces to the discrete 
case, where the behavior is well understood \cite{ruelle}. For non-zero 
but small $\epsilon$, the behavior does not depend on the precise value 
of $E_c$. In fact, starting with a recurrent configuration of the pile, 
and adding energy at some chosen site, we get exactly the same sequence 
of topplings for a range of values of $E_c$ \cite{redig}. To be precise, 
for any fixed initial configuration, and fixed driving sequence (of 
sites chosen for addition of energy), whether a site $x$ topples at time 
$t$ or not is independent of $E_c$, so long as we have $1 + \epsilon < 
E_c \le 2 -2 \epsilon$. In the following, we assume for simplicity that 
$E_c = 3/2$, and $0 \leq \epsilon \le 1/4$.

 It was shown in \cite{redig} that in this case, the stationary state has at 
most one site with energy $E(x,t)=0$ and all other sites have energy in the 
range $ 1 -\epsilon \le E(x,t) \le 1 +\epsilon$. The position of the empty site 
is equally distributed among all the lattice points. There are also some 
recurrent configurations in which all sites have energy $E(x,t) \geq 1 - \epsilon$.   
In such cases, we shall say that the site with zero energy is the site 
$L+1$.  Then, in the steady state, there is exactly one site with energy equal to $0$, and the 
$L+1$ different positions of the site are equally likely.

If $ E_c$ does not satisfy the inequality $ 1 + \epsilon  <  E_c \le
2 - 2 \epsilon$,  this simple characterization of the steady state is no 
longer valid. However, our treatment can be easily extended to those 
cases. Since the qualitative behavior of the model is the same in all 
cases, we restrict ourselves to the simplest case here.

It is easy to see that the toppling rules are in general not Abelian. 
For example, start with a two site model in configuration $(1.6,2.0)$ 
and $E_c=1.5$. The final configuration would be $(1.4, 0)$, or $(0, 
1.3)$,  depending on whether the first or the second site is toppled initially.  In our model, 
using the parallel update rule, the final configuration would be 
$( 1.0, 0.8)$.  A. Fey \textit{et al.} \cite{redig} have shown that only 
in one dimension, for $1+\epsilon < E_c$, the Zhang model has a restricted Abelian 
character, namely, that the final state does not depend on the order of topplings 
within an avalanche. However, topplings in two different avalanches do 
not commute.

\section{The propagator, and its  relation to the discrete Abelian model} 
It is useful to look at the Zhang model as a perturbation about the 
$\epsilon =0$ limit.  For sufficiently small $\epsilon$, given the site of 
addition and initial configuration,  the toppling sequence is {\it 
independent} of $\epsilon$.  It is also 
independent of the amount of energy of addition $\Delta_t$, and is same as 
the model with $\epsilon = 0$, which is the $1$-dimensional Abelian 
sandpile model with integer heights (hereafter referred to simply as 
ASM, without  further 
qualifiers).  
We decompose the energy variables as
\begin{equation}
\label{eq:1}
  E(x,t)=\mbox{Nint}[E(x,t)] + \epsilon \eta(x,t),
\end{equation}
where Nint refers to the nearest integer value. Then the integer part of 
the energy evolves as in the ASM. We write
\begin{equation}
\Delta_t = 1 + \epsilon u_t , \mathrm{~for ~~all~~}t.
\end{equation}
Here $u_t$ is uniformly distributed in the interval $[-1,+1]$. The 
linearity of energy transfer in toppling implies that the evolution of the 
variables $\eta(x,t)$ is independent of $\epsilon$. Thus, $\eta(x,t)$ is a 
linear function of $u_t$; the precise function depends on the sequence 
of topplings that took place. These are determined by the sequence of
addition sites $\{a_t\}$ up to the time $t$, and the initial configuration 
$C_0$. These together will be called the evolution history of the 
system up to time $t$, and denoted by  $\mathcal{H}_t$. We assume that at 
the starting time $t = 0$, the 
variables $\eta(x, t=0)$ are zero for all $x$, and the initial 
configuration is a recurrent configuration $C_0$ of the ASM. Then, from 
the linearity of the toppling rules, we can write $\eta(x,t)$ as a 
linear function of $\{u_{t'}\}$ for $1 \leq t' \le t$, and we can write for 
a given history $\mathcal{H}_t$,
\begin{equation}
  \label{eq:3}
  \eta(x,t|\{u_t\},\mathcal{H}_t)=  
  \sum_{t'=1}^{t}G(x,t|a_{t'},t',\mathcal{H}_t)u_{t'}.
\end{equation}
This defines the matrix elements $G(x,t|a_{t'},t',\mathcal{H}_t)$. These 
can be understood in terms of the probability distribution of the 
position of a marked grain in the ASM as follows. Consider the motion of 
a marked grain in the one dimensional height type BTW model. We start 
with configuration $C_0$ and add grains at sites according to the 
sequence $\{a_t\}$. All grains are identical except the one added at 
time $t'$, which is marked. In each toppling, the marked grain jumps to 
one of its two neighbors with equal probability. Consider the 
probability that the marked grain will be found at site $x$ after a 
sequence of relaxation processes at time $t$. We denote this probability as 
${\rm Prob}(x,t|a_{t'},t',\mathcal{H}_t)$. From the toppling rules in 
both the models, it is easy to see that 
\begin{equation}
 G(x,t|a_{t'},t',\mathcal{H}_t) = {\rm Prob}(x,t|a_{t'},t',\mathcal{H}_t).
\end{equation}
Averaging over different histories $\mathcal{H}_t$, we get the probability 
that a marked grain added at $x'=a_{t'}$ at time $t'$  is found at a 
position $x$ at time $t \ge t'$ in the steady state of the ASM.
Denoting the latter probability by $\mathrm{Prob_{ASM}}(x,t| x',t')$, we get
\begin{equation}
  \label{eq:5}
\overline{ G(x,t|x'=a_{t'},t',\mathcal{H}_t)} = 
\mathrm{Prob_{ASM}}(x,t|x',t'),
\end{equation}
where the over bar denotes averaging over  different 
histories $\mathcal{H}_t$, consistent with the specified constraints. 
Here, 
the constraint is that $\mathcal{H}_t$ must satisfy  $a_{t'} = x'$. At 
other places, the constraints may be different, and will be  specified  
if not clear from the context.

We shall denote the variance of a random variable $\xi$ by  $Var[\xi ]$.
From the definition in Eq. \eqref{eq:1}, it is easy to show that 
\begin{equation}
Var[E(x,t)] =   L / (L+1)^2 + \epsilon^2 Var[\eta(x,t)].
\end{equation}
Different $u_t$ are independent random variables, also independent of 
$\mathcal{H}_t$ and have zero mean. Let $Var[ u_t] = \sigma^2$. For the case 
when $u_t$ has a uniform distribution between $ -1$ and $ +1$, we have 
$\sigma^2  =  1/3$. Then, from Eq. \eqref{eq:3}, 
we get
\begin{equation}
  \label{eq:7}
  Var[\eta(x,t)] =  \sigma^2
\sum_{t'=1}^{t} \overline{G^2(x,t|a_{t'},t',\mathcal{H}_t)}.
\end{equation}
As $t \rightarrow \infty$, the system tends to a steady state, and the 
average in the 
right hand side of Eq. \eqref{eq:7} becomes a function of $t-t'$. Also, for a 
given $t'$, all values of $a_{t'}$ are equally likely. We define
\begin{equation}
  \label{eq:8}
F(x,\tau) \equiv \frac{1}{L} \lim_{t' \rightarrow \infty}\sum_{x'} 
\overline{G^2(x,t'+ \tau | 
x',t',\mathcal{H}_t)}.
\end{equation}
Then, for large $L$, in the steady state ($t$ large), the variance of energy at site 
$x$ is $1/L+\epsilon^2\Sigma^2(x)$, where 
\begin{equation}
  \label{eq:9}
  \Sigma^2(x) = \lim_{t \rightarrow \infty} Var[\eta(x,t)] = \sigma^2 \sum_{\tau=0}^{\infty} F(x,\tau).
\end{equation}
We define $\overline{\Sigma^2}$ to be the  average of $\Sigma^2(x)$ over 
$x$.
\begin{equation}
\overline{\Sigma^2} = \frac{1}{L} \sum_x \Sigma^2(x).
\end{equation}
Evaluation of $G(x,t|x',t',\mathcal{H}_t)$ for a given history 
$\mathcal{H}_t$ and averaging over $\mathcal{H}_t$ is quite tedious for 
$t > 1$ or $2$.  For $\overline{G}$, the problem has been studied in 
the context of residence times of grains in sand piles, and some exact results 
are known in specific cases \cite{punyabrata}.  For $\overline{G^2}$, 
the calculations are much more difficult.  However, some simplifications 
occur in large $L$ limit. We discuss these in the next section.
\section{Calculation of $\Sigma^2(x)$ in large-$L$ limit}
In order to find the quantity 
$F(x,\tau)$ in Eq. (8), we have to 
average $ G^2(x,t | x',t',\mathcal{H}_t)$ over all possible histories 
$\mathcal{H}_t$, which is quite difficult to evaluate exactly. 
However, we can determine the leading behavior of $F(x,\tau)$ in this limit. 

We use the fact that the path of a marked grain in the ASM is a random 
walk \cite{punyabrata}.  Consider a particle that starts away from the 
boundaries, at $x'= \xi L$, with $L$ large, and $0 < \xi < 1$. If it 
undergoes $r(\mathcal{H}_t)$ topplings between the time $t'$ and $t = t' + 
\tau$ under 
some particular history $\mathcal{H}_t$, then its probability distribution 
is approximately a Gaussian, centered at $x'$ with width $\sqrt{r}$.  
Then, we have 
\begin{equation} 
G(x,t|x',t', \mathcal{H}_t) \simeq  \frac{1}{\sqrt{2 \pi r(\mathcal{H}_t) }} \exp\left( - 
\frac{( x - x')^2} {2 r(\mathcal{H}_t)}\right). 
\end{equation}
Using this approximation for $G$, summing over $x'$, we get
\begin{equation}
  \label{eq:12}
\sum_{x'} G^2(x,t|x',t',\mathcal{H}_t) \simeq \frac{1}{2\sqrt{\pi 
r(\mathcal{H}_t)}}.
\end{equation}
Thus, we have to calculate the average of $1/\sqrt{r(\mathcal{H}_t)}$ over 
different histories. Here $r(\mathcal{H}_t)$ was defined as the number of 
topplings undergone by the marked grain. Different possible 
trajectories of a marked grain, for a given history, do not have the 
same number of topplings. However, if the typical displacement of the 
grain is much smaller than its distance from the end, differences 
between these are small, and can be neglected. There are typically 
${\cal O}(L)$ topplings per grain per avalanche in the model, and a grain moves a 
typical distance of ${\cal O}(\sqrt{L})$ in one avalanche. Then, we can 
approximate $r(\mathcal{H}_t)$ by $N(x')$, the number of topplings at $x'$.

 Let the number of topplings at $x'$ at time steps $\tau = 0, 1 , 2, \ldots$ be 
denoted by $N_0, N_1, N_2, \ldots$. Then, $N(x') = N_0 + N_1 + N_2 + \cdots$.  
It can be shown 
that the number of topplings in 
different avalanches in the one dimensional ASM are nearly uncorrelated (In 
fact the correlation function between $N_i$ and $N_j$ varies as 
$(1/L)^{|i -j|}$.).  By 
the central limit theorem for sum of weakly correlated random variables, 
the  mean value of $N$ grows linearly with $\tau$, but the standard 
deviation increases only as $\sqrt{\tau}$. Then, for $\tau \gg 0$, the 
distribution is sharply peaked about the mean, and $\langle1/\sqrt{N} \rangle \simeq 
1/\sqrt{\langle N \rangle}$. 

Clearly, for $\tau \gg 0$,  $\langle N \rangle = \tau 
\bar{n}(x')$, where $\bar{n}(x')$ is the mean number of topplings per 
avalanche at $x'$ in the ASM,  given by
\begin{equation}
\bar{n}(x = \xi L)= L \xi  (1 - \xi)/2.
\end{equation}
The upper limit on $\tau$  for the validity of the above argument 
comes from the 
requirement that the width of the Gaussian be much less than the 
distance from the boundary, (without any loss of generality, we can assume 
that $\xi < 1/2$, so that it is the left boundary ), else we cannot neglect 
events where the marked grain 
leaves the pile. This gives  
$\sqrt{\tau \bar{n}(x)} \ll \xi L$, or equivalently, $\tau \ll \xi L. $
Thus we get,
\begin{equation}
  F(x,\tau) \simeq \frac{C_1}{L} [ \tau L \xi (1 -\xi)]^{-1/2}, \mathrm{~ for~~} 0 \ll \tau 
\ll \xi L,
\end{equation}
where $C_1$ is some constant.

Also, we know that for $\tau \gg L$, the probability that the grain 
stays in the pile decays exponentially as $\exp( -\tau/L)$ \cite{punyabrata}.
Thus, $\overline{G}$, and also $\overline{G^2}$ will decay exponentially 
with $\tau$, for $\tau \gg L$. Thus, we have, for some constants $C_2$ and 
$a$,
\begin{equation}
  F(x,\tau) \simeq \frac{C_2}{L^2} \exp( - a \tau/L), \mathrm{~~for~} \tau \gg L.
\end{equation}
It only remains to determine the behavior of $ F(x,\tau)$, for $\xi L 
\ll \tau \ll L$.   In this 
case, in the ASM, there is a significant probability that the marked 
grain leaves 
the pile from the end. This results in a faster decay of $G$, and hence 
of $F$ with time.   We argue below that the behavior of the function 
$F(x,\tau)$ is 
given by
\begin{equation}
  F(x,\tau) \sim \frac{C_3}{L\tau}, \mathrm{~~for~~} \xi L  \ll \tau \ll L,
\end{equation}
where $C_3$ is some constant.
 This can be seen as follows:  Let us consider 
the special case when the particle starts at a site close to the boundary. 
Then $\bar{n}(x)$ is approximately a linear function of $x$ 
for small $x$. Its spatial variation cannot be neglected, and Eq. \eqref{eq:12} is 
no longer valid.  We 
will now argue that  in this case
\begin{equation}
  \overline{G(x,t' + \tau |x', t')} \simeq x' \tau^{-2} \exp( -x/ \tau), \mathrm{~~for~~} 
0  \ll \tau \ll L.
\end{equation}
The time evolution of $Prob_{ASM}(x,t|x',t')$ in Eq. \eqref{eq:5} is well described as a diffusion
with diffusion coefficient proportional to $\bar{n}(x)$ which is the mean
number of topplings per avalanche at $x$ in the ASM \cite{punyabrata}.
For understanding the long-time survival probability in this problem, we 
can equivalently  consider  the problem in a continuous-time version: 
consider a  random walk on 
a half line where sites are labelled by positive integers, and the jump 
rate out of a site $x$ is proportional to $x$. A particle starts at site 
$x =x_0$ at time $t=0$. If $P_j(t)$ is the probability that the particle is 
at $j$ at time $t$, then the equations for the time-evolution of $P_j(t)$ 
are, for all $ j>0$,
\begin{equation}
  \label{eq:18}
\frac{d}{dt} P_j(t) = (j+1) P_{j+1}(t) +(j-1) P_{j-1}(t) - 2j 
P_j(t).
\end{equation}
The long time solution starting with $P_j(0) = \delta_{j,x_0}$ is 
\begin{equation}
  P_j(t) \simeq x_0 t^{-2} \exp(- j/t)
\end{equation}
for $t \gg x_0$ and large $j$. The 
probability that the particle survives till time $t$ decreases as $1/t$ 
for large $t$. We have discussed the calculation in the Appendix. 

Using Eq. \eqref{eq:5}, we see that $ \overline{G(j,t'+\tau|x_0,t')}$ scales as $ 
x_0/\tau^2 $.  It seems reasonable to assume that $\overline{G^2}$ will 
scale as $\overline{G}^2$. Then, each term in the 
summation for $F(x,\tau)$ in Eq. \eqref{eq:8} scales as 
$x_0^2/\tau^4$, and there are $\tau$ such terms, as the  sum over $x_0$ has an 
upper cutoff proportional to $\tau$, and so $F(x,\tau)$ varies as 
$1/\tau$ for $ L \gg \tau \gg x_0$. This concludes the argument.

We can put these three limiting behaviors into a single functional form 
that interpolates between these, as
\begin{equation}
  F(x,\tau) \simeq \frac{1}{L} \frac{K \exp( - a \tau/L)}{\tau + B \sqrt{\tau L \xi (1 
-\xi)}},
\end{equation}
where $K$, $a$ and $B$ are some constants.
In Section V, we will see that results from numerical simulation 
are consistent with this phenomenological 
expression.  
 
Using this interpolation form in Eq. \eqref{eq:9}, and converting the sum 
over $\tau$ to an integration over a variable $u = \tau/L$,  we can write
\begin{equation}
\Sigma^2(x = \xi L) \simeq \frac{\sigma^2}{L} \int_0^{\infty}du \frac{ K 
\exp( - a u)}{u + B 
\sqrt{ u \xi ( 1 - \xi)}}.
\end{equation}
This integral can be simplified by a change of variable $ au = z^2$, 
giving
\begin{equation}
\Sigma^2(x = \xi L) \simeq \frac{ K \sigma^2}{L} I\left(B'\sqrt{\xi(1 - \xi)}\right),
\end{equation}
where $K, B'$ are  constants, and $I(y)$ is a function defined by
\begin{equation}
I(y) = 2\int_0^{\infty} dz \frac{ \exp( - z^2)}{ z + y}.
\end{equation}
It is easy to verify that $I(y)$ diverges as $\log ( 1/y)$ for small 
$y$.  In 
particular, we note that the exponential term in the integral 
expression for $I(y)$ has a significant contribution only for $z$ near 
$1$.  We may approximate this by dropping the exponential factor, and 
changing the upper limit of the integral to $1$.  The resulting integral 
is easily done, giving
\begin{equation}
  \label{eq:24}
\Sigma^2(x=\xi L) \simeq \frac{ K' \sigma^2}{ L} \log \left( 1 + 
\frac{1}{B'\sqrt{\xi(1 -\xi)}}\right),
\end{equation} 
where $K'$ is some constant.
Averaging $\Sigma^2(x)$ over $x$, we get a behavior $ 
\overline{\Sigma^2}(x) \simeq 1/L$. Of course, the answer is not exact, and one could have constructed 
other interpolation forms that have the same asymptotic behavior.
We will see in the next Section that results from numerical simulations
for $\Sigma^2(x)$ can be fitted very well to the phenomenological
expression in Eq. \eqref{eq:24}.
\section{Numerical results}
\begin{figure}
\begin{center}
\includegraphics[scale=0.52,angle=270]{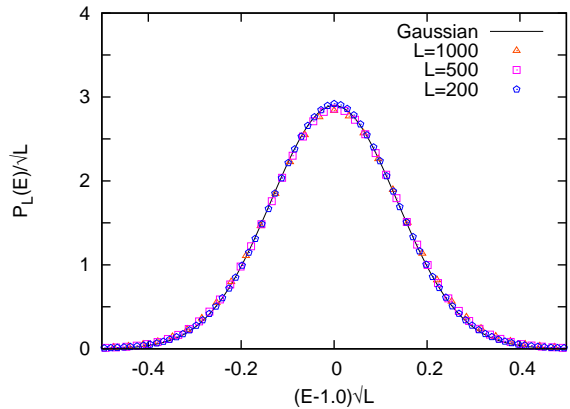}
\caption{(Color online) Scaling collapse of the probability distribution
$\mathcal{P}_L(E)$ of energy per site in the steady state for different systems 
of size $200$, $500$ and $1000$. The distribution is well described by a Gaussian of width $0.136$.}
\label{patch1}
\end{center}
\end{figure}

We have tested  our non-rigorous  theoretical arguments against results 
obtained from numerical simulations. In Fig. \ref{patch1}, we have plotted the 
probability distribution $\mathcal{P}_L(E)$ of energy at a site, averaged over all sites. We used $L= 
200$, $500$ and $1000$, and averaged over $10^8$ different configurations 
in the steady state. We plot the scaled distribution function 
$ \mathcal{P}_L(E)/\sqrt{L}$ versus the scaled energy $( E - \bar{E}) 
\sqrt{L}$. A good collapse is seen, which verifies the fact that the width 
of the peak varies as ${L^{-1/2}}$. 

\begin{figure}
\begin{center}
\includegraphics[scale=0.65,angle=270]{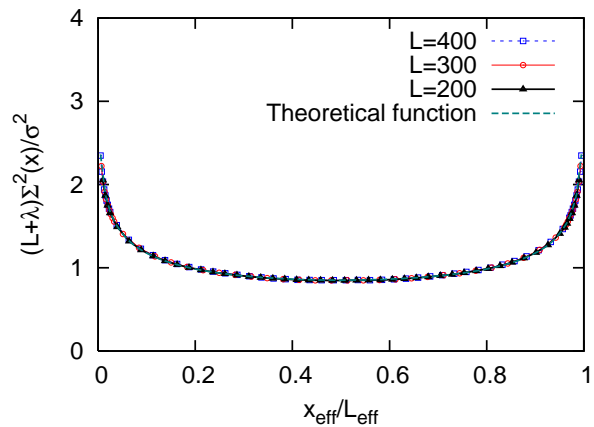}
\caption{(Color online) Scaling  collapse of $\Sigma^2(x)/\sigma^2$ at site $x$ 
for systems of different length $L$. }
\label{patch2}
\end{center}
\end{figure}

The  dependence of the variance of $E(x,t)$ on
$x$ is plotted in Fig. \ref{patch2} for systems of length $200$, $300$ and 
$400$. The data was  obtained by averaging over $10^8$ 
avalanches. We plot $( L + \lambda) \Sigma^2(x)/\sigma^2$ versus 
$x_{eff}/L_{eff}$, where $x_{eff}$ differs from $x$ by an amount 
$\delta$ to take into account the corrections due to end effects. Then, for 
consistency, $L$ is replaced by $L_{eff} = L + 2 \delta$.   For 
the specific choice of $\lambda=5 \pm 1$ and $\delta=1.0 \pm 0.2$, we get 
a good collapse of the curves for different $L$.  We also show a fit to 
the proposed interpolation form in Eq. \eqref{eq:24}, with $K'=1.00 \pm
0.01$ and $B'= 1.5 \pm 0.2$. We see that the fit is very good.

\begin{figure}
\begin{center}
\includegraphics[scale=0.65,angle=270]{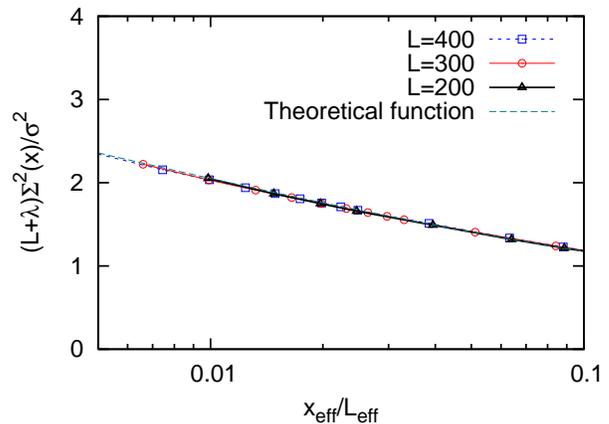}
\caption{(Color online) The same plot in Fig. \ref{patch2} resolved more at the  left boundary of 
the model and taking $x$ axis  in log scale.}
\label{patch3}
\end{center}
\end{figure}

In order to check the logarithmic dependence of $\Sigma^2(x)$
on $x$ for small $x$, we re-plot the data in Fig. \ref{patch3} using 
logarithmic scale for $x$.  We get a good collapse of the data for 
different $L$, supporting our proposed dependence in Eq. \eqref{eq:24}.
\section{Concluding remarks}
To summarize, we have studied the emergence of quasi-units in the
one-dimensional Zhang sandpile model. The variance of energy variables 
in the steady state is governed by the balance between two competing 
processes. The randomness in the drive i.e., the energy of addition, 
tends to increase the variance in time. On the other hand, the topplings 
of energy variables tend to equalize the excess energy by distributing 
it to the nearby sites. There are on an average $O(L^2)$ topplings per 
avalanche. Hence, in one dimension there are, on an average, $O(L)$ 
topplings per site per avalanche. For large system size, the second 
process dominates over the first and the variance becomes low. We have 
shown that the variance vanishes as $1/L$ with increasing system size 
and the probability distribution of energy concentrates around a 
non-random value which depends on the energy of addition. We have also 
proposed a functional form for the spatial dependence of variance of 
energy which incorporates the correct limiting behaviors, and matches 
very well with the numerical data.

An interesting question is whether one can extend these arguments to the 
two-dimensional Zhang model. In this case, there are several peaks in 
the distribution of energies at a site, but there are some numerical 
evidences for the sharpening of the peaks as the system size is increased. 
However, as the number of topplings per site varies only as $\log L$,
the width is expected to decrease much more 
slowly with $L$, and the fluctuation effects can be much stronger. This 
remains an open question for further study.

\appendix*
\section{}
Here we discuss the solution of the Eq. \eqref{eq:18} for the starting values 
given by  
\begin{equation}
  \label{eq:A1}
  P_j(t =0)=\alpha^{j-1}.
\end{equation}
We start with an ansatz $P_j(t)=b_t\exp(-a_tj)$, where both $a_t$ and $b_t$ are functions only of 
$t$. This form satisfies the Eq. \eqref{eq:18} for all $j$,
$t>0$, if $a_t$ and $b_t$ satisfy
\begin{eqnarray}
  \label{eq:A2}
  \frac{da_t}{dt}&=& 2-e^{a_t}-e^{-a_t}, \\
  \label{eq:A3}
  \frac{db_t}{dt}&=& b_t (e^{-a_t}-e^{a_t}).
\end{eqnarray}
To solve the Eq. \eqref{eq:A2}, we first make a change of variable $z=e^{-a_t}$. 
In terms of $z$, the equation becomes $dz/dt=(1-z)^2$, which can be easily integrated
to give
\begin{equation}
  \label{eq:A4}
  e^{-a_t}=\frac{t+A-1}{t+A},
\end{equation}
where $A$ is an integration constant. To satisfy the
initial condition in Eq. \eqref{eq:A1}, we choose 
\begin{equation}
  A=(1-\alpha)^{-1}.
\end{equation}
Similarly, to solve the equation for $b_t$, we use the form of $e^{-a_t}$ given in 
Eq. (A4) and get
\begin{equation}
  \frac{db_t}{dt}= b_t\frac{1-2(t+A)}{(t+A)(t+A-1)}.
\end{equation}
This can be integrated to give
\begin{equation}
  b_t=\frac{B}{(t+A)(t+A-1)},
\end{equation}
\\
where $B$ is an integration constant. 
Then the probability can be written as 
\begin{equation}
  P_j(t)=B\frac{(t+A-1)^{j-1}}{(t+A)^{j+1}}.
\end{equation}
To satisfy the initial condition at $t=0$,
we choose the integration constant $B=(1-\alpha)^{-2}$.
Then, with these values of $A$ and $B$, we have the solution for all $j$, $t>0$, given by
\begin{eqnarray}
  \label{eq:A9}
  P_j(t)&=&\frac{\left[(1 -\alpha) t+\alpha \right]^{j-1}}{\left[ ( 1-\alpha) t+ 1 \right]^{j+1}}\nonumber\\
  &=&\phi_j(\alpha,t),\mathrm{~ ~ ~ ~ ~ say}.
\end{eqnarray}
Now, as $\phi_j(\alpha,t)$ satisfies the Eq. \eqref{eq:18},
\begin{equation}
  \psi_{j,n}(\alpha,t)=\frac{1}{(n-1)!}\frac{\partial^{n-1}\phi_j(t)}{\partial\alpha^{n-1}}
\end{equation}
will also satisfy the equation for any natural number $n$. In addition,
\begin{equation}
  \psi_{j,n}(\alpha=0,t=0)=\delta_{j,n}.
\end{equation}
Hence, we see that the solution of the Eq. \eqref{eq:18}, starting
with $P_j(t)=\delta_{j,n}$ at $t=0$ is
\begin{equation}
P_j(t)=\psi_{j,{n}}(\alpha=0,t)=\frac{1}{(n-1)!}\frac{\partial^{n-1}\phi_j(\alpha,t)}{\partial\alpha^{n-1}}\mid_{\alpha=0},
\end{equation}
for all $j$, $t>0$, where $\phi_j(\alpha,t)$ is given in Eq. \eqref{eq:A9} and 
$n$ is any natural number.

It can be shown that for large $t$ and $j$, the solution
asymptotically becomes $P_j(t)= n t^{-2}\exp(-j/t)$.

\end{document}